\documentclass[journal]{IEEEtran}
\usepackage{graphicx}
\usepackage{authblk}
\usepackage{tikz, amsmath, amssymb, bm, color}
\usepackage{hyperref}
\usepackage{soul,color}
\usepackage{siunitx}
\usepackage{csquotes}
\usepackage{enumitem}
\usepackage{subfig}
\begin{document}

\title{Towards Enabling \textit{High-Five} Over WiFi}

\author[$\dag$]{Vineet Gokhale}
\author[$\ddag$]{Mohamad Eid}
\author[$\dag$]{Kees Kroep}
\author[$\dag$]{R. Venkatesha Prasad}
\author[$\star$]{Vijay Rao}

\affil[$\dag$]{Delft University of Technology, The Netherlands}
\affil[$\ddag$]{New York University Abu Dhabi, UAE}
\affil[$\star$]{Cognizant Technology Solutions, The Netherlands}

\maketitle

\begin{abstract}
The next frontier for immersive applications is enabling sentience over the Internet. \textit{Tactile Internet (TI)} envisages transporting skills by providing Ultra-Low Latency (ULL) communications for transporting \textit{touch senses}. 
In this work, we focus our study on the first/last mile communication, where the future generation WiFi-7 is pitched as the front-runner for ULL applications. We discuss a few candidate features of WiFi-7 and highlight its major pitfalls with respect to ULL communication. 
Further, through a specific implementation of WiFi-7
(\textit{vanilla WiFi-7}) in our custom simulator, we demonstrate the impact of one of the pitfalls -- standard practice of
using jitter buffer in conjunction with frame aggregation --
on TI communication. To circumvent this, we propose \textit{\underline{No}n-\underline{Bu}ffered \underline{S}cheme (NoBuS)} -- a simple MAC layer enhancement for enabling TI applications on WiFi-7. NoBuS trades off packet loss for latency enabling swift synchronization between the master and controlled domains. 
Our findings reveal that employing NoBuS yields a significant improvement in RMSE of TI signals. 
Further, we show that the worst-case WiFi latency with NoBuS is \SI{3.72}{ms} -- an order of magnitude lower than vanilla WiFi-7 even under highly congested network conditions.
\end{abstract}


\section{Introduction}
\label{sec:intro}

The Covid-19 pandemic has dramatically increased our reliance on interactive audio-video teleconferencing.
The disruptive vision of \textit{Tactile Internet (TI)}  takes this interactivity a step further by enabling transportation of physical actions coupled with \textit{haptic} (touch) feedback over long distances~\cite{fettweis2014tactile}. This facilitates exchange of touch-based gestures, such as \lq\textit{High-Fives}\rq, hugs, and handshakes, revolutionizing our ability to be sentient over the Internet.  

Such an immersive experience enables seamless \textit{skill transfer}, manifesting the potential to revolutionize many fronts of human lives. TI will lay a strong foundation for accelerating the progress of the \textit{Industry 4.0} revolution that enables industrial processes to be remotely executed through controlling telerobots. Critical healthcare services can be democratized through telesurgery; surgeons can conduct medical procedures over long distances with the same precision and speed as a conventional surgery. TI will also significantly redefine several other sectors such as education, entertainment, e-commerce, and disaster management.

\begin{figure*}[!t]
    \centering
    \includegraphics[width=0.8\textwidth]{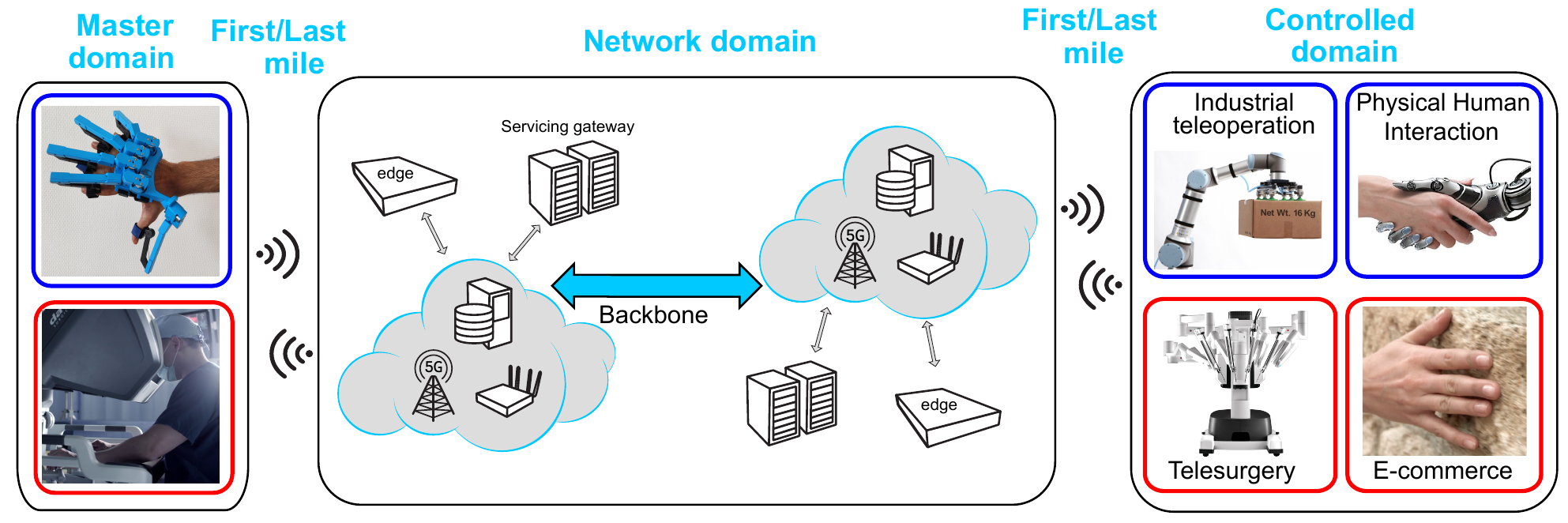}
    \caption{Schematic representation of TI depicting the master, network, and controlled domains as well as a few popular use-cases.}
    \label{fig:TI_arch}
\end{figure*}

Fig.~\ref{fig:TI_arch} depicts a high-level representation of various domains and use-cases of TI. The human operator and the teleoperator in the \textit{master} and \textit{controlled domains}, respectively, are equipped with sensor and actuator devices.
Such devices include robotic arms, exoskeletons, and haptic suits. The motion commands (position, velocity, and torque) generated by the operator's actions are communicated over the \textit{first mile} to the network domain that facilitates their fast communication. These commands are then delivered over the \textit{last mile} to the controlled domain where they are utilized by the teleoperator to replicate operator's actions. The dynamics of the controlled domain, captured as force, audio, and video, are fed back to the operator.

The promising benefits of TI accompany several challenges in the context of sensing, actuation, and communication. The foremost challenge with respect to communication is the ultra-low latency (ULL) requirement of round-trip latency of sub-\SI{10}{ms}~\cite{holland2019}. High quality of teleoperation can only be maintained if the haptic feedback corresponding to an action is provided to the operator within the ULL deadline. 
While innovations in 5G are making giant strides towards meeting ULL demands, designing specialized solutions for TI would further accelerate the progress towards realizing seamless teleoperation.
Additionally, the TI community also recommends an ultra-reliability requirement of up to \SI{99.9999}{\%}~\cite{holland2019}. However, unlike the ULL requirement, which is experimentally quantified, the ultra-reliability requirement has no reasonable substantiation and is only a speculation of the necessary performance guarantee. In fact, a recent work ~\cite{etvo} that involved rigorous subjective experiments reveals that even a significant amount (nearly \SI{50}{\%}) of packet losses barely introduces any disturbance to the users. This indicates the possibility of ultra-reliability being an overkill for at least some TI applications that have the same characteristics as the TI task investigated in work~\cite{etvo}. Extending this observation to a broader range of TI applications necessitates further systematic studies. Until the TI reliability requirement is well-established, treating ultra-reliability as a key requirement may lead to over-designing TI solutions.
Hence, in this work, we focus on the ULL requirement of TI. 
Note that in this work, we provide objective
measurements that align with the claims on reliability requirement made in~\cite{etvo}. This further justifies our decision to focus on ULL requirement only.

Satisfying TI's ULL constraint demands significant advancements at every hop in the network including the network domain and  first/last mile. In this work, we focus on the first/last mile communication that presents several interesting research opportunities. The ubiquity of WiFi makes it one of the front-runners in the first/last mile communication.
Further, extensive efforts are underway to enable ULL applications through the future generation \textit{WiFi-}7 (IEEE 802.11be).
A few works, such as \cite{naik,adame}, have conducted simulation-based studies to provide a preliminary understanding of the latency performance of some candidate features of WiFi-7. However, these works are generic and do not consider the specifics of TI applications.      

In this work, we take the first step towards answering the following questions -- (i) How well are the candidate features of the future WiFi-7 suited for TI applications? (ii) How to overcome their limitations?

Our contributions in this work are the following. First, we provide an overview of a few candidate features of WiFi-7. We  highlight major pitfalls of WiFi-7 that might constrain its usage for TI communications. 
Secondly, we investigate one of the pitfalls by considering a specific implementation of WiFi-7 involving the \SI{320}{MHz} bandwidth candidate feature, Orthogonal Frequency Division Multiple Access (OFDMA), and frame aggregation in our open-sourced custom MAC simulator. We call this implementation as \textit{vanilla WiFi-7} and take it as the baseline for our performance evaluations.
By using realistic TI data traces, we objectively demonstrate the severe limitations of vanilla WiFi-7 on the quality of teleoperation. 
To address this issue, we propose \textit{\underline{No}n-\underline{Bu}ffered \underline{S}cheme (NoBuS)}  -- a MAC layer enhancement tailored for TI applications to facilitate swift synchronization between the master and controlled domains. Our experiments demonstrate that NoBuS can enable TI communication with a worst-case WiFi latency of approximately \SI{3.72}{ms} -- an order of magnitude lower than that of vanilla WiFi-7. Further, we demonstrate that employing NoBuS yields a significant reduction in RMSE of the reconstructed signal compared to vanilla WiFi-7.

\section{A WiFi Primer}
\label{sec:primer}
For providing a concrete idea of WiFi developments, we discuss some existing features of WiFi systems crucial for understanding our work (Sec.~II-A) and also discuss candidate features~\cite{khorov} that are expected to play key roles in driving WiFi's ULL objectives (Sec.~II-B). Further, we present some of the potential pitfalls of WiFi-7 (Sec.~II-C).

\subsection{Existing features of WiFi}
\label{subsec:existing}
\subsubsection{Frame aggregation}
\label{subsub:agg}
This is a key bandwidth-saving mechanism of WiFi introduced in 802.11n. 
If a device's (access point (AP) or station (STA)) data generation rate exceeds its transmission rate, the MAC layer aggregates the MAC Protocol Data Units (MPDUs) into an aggregated MPDU (AMPDU). AMPDUs are transmitted once the wireless medium becomes accessible. Transmitting an AMPDU instead of separate MPDUs leads to a substantial reduction in the protocol header overhead as well as the medium access latency.

\begin{figure*}[!t]
    \centering
    \includegraphics[width=0.8\textwidth]{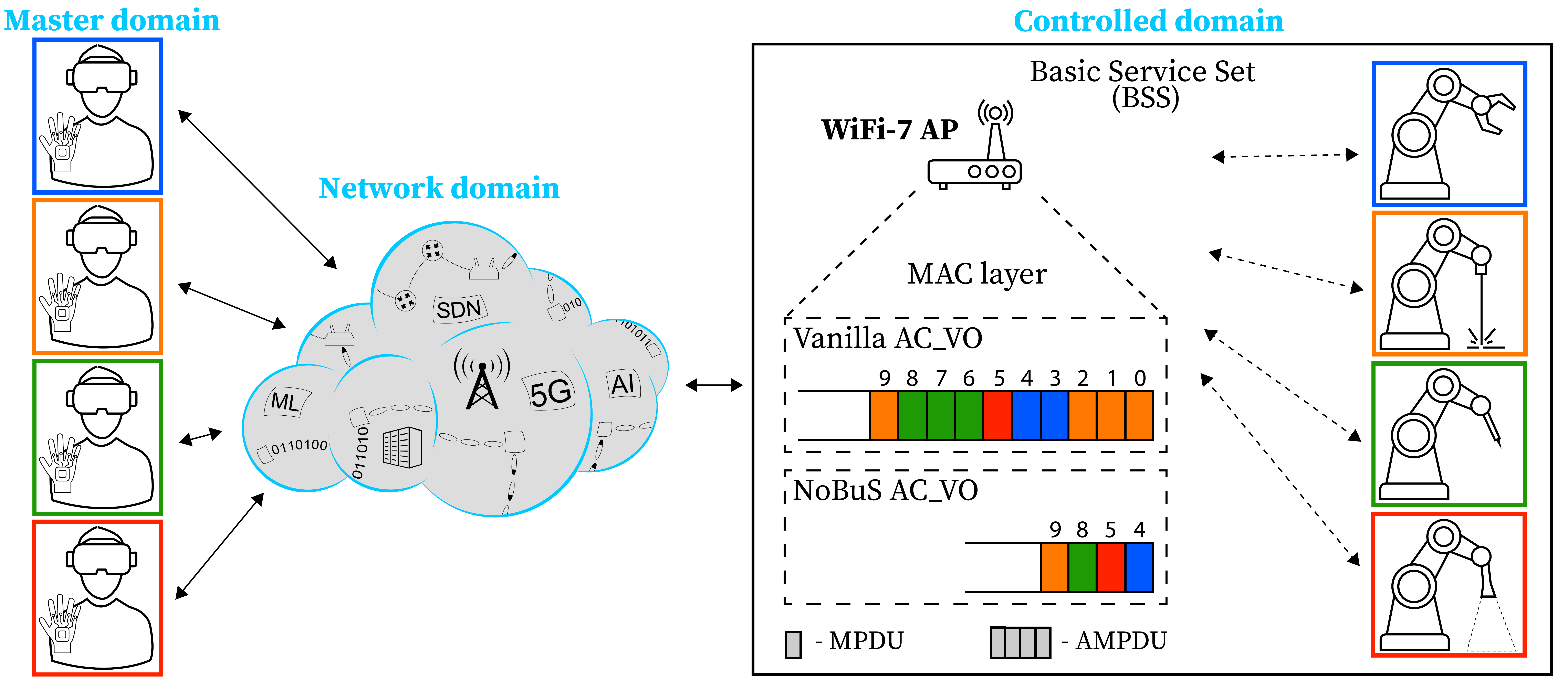}
    \caption{Schematic representation of a generic, WiFi-7 enabled TI setup used in this work. The MAC transmit queue of AP is depicted to contrast the frame aggregation of vanilla WiFi-7 and \textit{non-buffering} principle of NoBuS.}
    \label{fig:factory}
\end{figure*}

\subsubsection{OFDMA} 
This is a multiple access feature of WiFi introduced in 802.11ax (WiFi-6)~\cite{ref:80211ax}. Unlike in legacy WiFi systems (pre WiFi-6), where a STA utilizes the entire channel bandwidth during the medium access (single-user mode), OFDMA enables simultaneous medium access by multiple STAs (multi-user mode)~\cite{naik}. 
When a STA wins the contention, it employs the single-user mode for AMPDU transmissions, akin to legacy WiFi. OFDMA is triggered only when AP wins the contention by dividing the channel bandwidth into several resource units (RUs). OFDMA is employed on both uplink (UL) and downlink (DL). 

\vspace{0.6mm}
\noindent\textul{DL-OFDMA}: 
The AP employs DL-OFDMA to simultaneously transmit to multiple STAs by assigning a dedicated RU per STA. This assignment is communicated with the STAs so they can restrict listening to their own RUs. 

\vspace{0.6mm}
\noindent\textul{UL-OFDMA}:
The AP also provisions UL transmissions during its medium access through UL-OFDMA. 
In a simplistic implementation, the AP could trigger UL-OFDMA once it completes DL-OFDMA. The AP allows for UL-OFDMA using two modes -- scheduled access (SA) and random access (RA). In the SA mode, the AP assigns RUs based on the transmit queue of STAs to conduct collision-free UL transmission. In the RA mode, the RU access by the STA is random based on a backoff counter. The AP can employ the two modes in conjunction, in which case a fraction of the total RUs is dedicated for SA. The RA STAs contend for the remaining RUs.

\subsection{Candidate features of WiFi-7}
\label{subsec:candidate}
\subsubsection{PHY layer enhancements} To enhance multi-user, multiple-input, multiple-output (MU-MIMO) capacity, WiFi-7 proposes to increase the number of spatial streams from 8 to 16. Additionally, the maximum bandwidth will be doubled from 160 to 320~MHz. Furthermore, a higher modulation and coding scheme, 4096-QAM, will likely be included in the standards.

\subsubsection{Multi-Link Operation (MLO)} As opposed to previous versions of WiFi, MLO enables WiFi-7 devices to simultaneously contend across multiple frequency bands, namely 2.4, 5, and 6\,GHz. This allows them to access the band that becomes available at the earliest. 
Preliminary investigations show MLO's potential to reduce the worst-case medium access latency by an order of magnitude compared to single-link approach~\cite{naik}.  

\subsubsection{Separate AC for ULL streams}
\label{subsec:newac}
Currently, WiFi provides a different access categories (AC) for audio and video streams, with voice having the highest priority. 
Some WiFi-7 submissions are proposing to introduce a new AC for ULL streams with a priority higher than voice's~\cite{kim}. This is expected to significantly reduce the medium access latency. 

\subsubsection{Preemptive scheduling} 
\label{subsec:preempt}
Longer MAC frames, such as those carrying video samples, can hold up subsequent frame transmissions from other devices leading to higher latency. While this is acceptable for conventional voice and video applications, it could be catastrophic for TI applications due to their ULL constraint. To circumvent this issue,
WiFi-7 candidate solutions propose to pause an ongoing transmission of a lower priority frame in order to pave way for transmission of a higher priority frame when generated.

\subsection{WiFi-7 Pitfalls}
\label{subsec:pitfall}
Despite the above proposed enhancements, we identify three major inadequacies of WiFi-7 in supporting TI applications. 

\subsubsection{Standard practice causes sluggishness}
\label{subsubsec:standard}
Typically, humans are much more sensitive to audio-video loss than haptic loss. A loss of up to \SI{1}{\%} for conventional audio-video communication is acceptable~\cite{marshall}, whereas even up to \SI{90}{\%} haptic loss can go unnoticed \cite{steinbach}. 
On the contrary, if the haptic feedback is not provided within ULL limits, human operators cannot continue to teleoperate effectively \cite{holland2019}. Hence, haptic feedback is \textit{latency-sensitive}.

As a standard practice, WiFi utilizes frame aggregation (Sec.~\ref{subsub:agg}) to serve loss-sensitive applications as it aims to maximize the number of samples delivered. A jitter buffer is employed at the receiver's application layer to cope with this bursty arrival of samples. This increases the waiting time of the samples (prior to display). 
This higher latency could result in WiFi-7 not supporting turnkey use-cases such as safety-critical TI applications. Although this approach was designed to boost the performance of conventional audio-video applications, it is a \textit{shot-in-the-foot} for TI applications.

\subsubsection{Haptics AC}
\label{subsub:inter}
While using a separate AC for haptic stream seems to be an addition of a highly-desired feature, it creates an undesired consequence: TI applications are characterized by a heavy temporal correlation between voice, video, and haptic streams. Hence, maintaining inter-stream synchronization is crucial during media display. Therefore, the medium access should also be driven by the constraints for inter-stream synchronization, the lack of which may heavily obstruct TI applications over WiFi-7.

\subsubsection{Preemptive scheduling}
\label{subsub:preempt}
This feature, although beneficial for ULL applications, needs a thorough investigation, especially when employed in conjunction with a new AC. Video frames may be continually preempted for haptic frames causing non-compliance to video latency deadlines. Therefore, such context-unaware preemptive scheduling may impede TI sessions. 

The standard practice discussed in Sec.~\ref{subsubsec:standard} will undoubtedly be a part of the WiFi-7 standards as it is based on existing WiFi features. As we show later in Sec.~\ref{subsec:perf}, it introduces a few tens of milliseconds of additional latency. Hence, in this work, we take an important step towards addressing this issue.

\section{WiFi-7 Performance Assessment}
\label{sec:ti-wifi}

\subsection{Tactile Internet Setup}
\label{subsec:usecase}
The TI setup considered in this work is generic across several use cases, such as telesurgery and a connected factory. 
Let us take the example of a connected factory. It consists of manufacturing processes remotely executed by operators via controlling teleoperators located inside the factory as shown in Fig.~\ref{fig:factory}. 
For simplicity, we consider a scenario where the controlled domain is WiFi-7 enabled and each operator is controlling a separate teleoperator (the pairs are color-coded). 
Each WiFi-7 Basic Service Set (BSS) consists of an AP and multiple STAs (teleoperators). Since we focus on studying the intra-BSS dynamics, we depict a single BSS in  Fig.~\ref{fig:factory}. 
Since our focus is only on the WiFi network, we assume that other network links are uncongested and the WiFi network serves only TI streams. Background traffic, like loss-based TCP, is known to be detrimental for TI applications~\cite{ref:TI_tcp}. Therefore, it is important to carry out TI communication in tightly-controlled WiFi networks.

\begin{table}[!t]
\small
\centering
\resizebox{0.3\textwidth}{!}{
 \begin{tabular}{|c|c|} 
 \hline
 \textbf{Parameter} & \textbf{Value} \\
  \hline
 Frequency band & 5GHz\\
 \hline
 Channel bandwidth & 320MHz\\
\hline
 Subcarrier spacing & 78.125kHz \\
 \hline
 Access category & Voice (AC\_{VO}) \\ 
 \hline
 Min. contention window & 3 \\
 \hline
 Max. contention window & 7 \\
 \hline
 Retransmission limit & 7 \\
 \hline
 MCS index & 9 \\
 \hline
 Slot size & 9$\mu$s \\
 \hline
 AIFS & 34$\mu$s \\
 \hline
 SIFS & 16$\mu$s \\
 \hline
 Max. PPDU duration & 5.4ms \\
  \hline
 Guard interval & 0.8$\mu$s \\
 \hline
 Aggregation type & AMPDU \\
 \hline
\end{tabular}
}
\caption{WiFi-7 configuration parameters used in our simulations.}
\label{tab:wifi}
\end{table}

\subsection{Network Simulations}
\label{subsec:wifi}

\subsubsection{The Simulator}
Since this is a first-of-its-kind work investigating TI communication over WiFi-7, we resort to a simulation-based study. 
At the time of writing this paper, NS-3 has no complete OFDMA implementation in its official release.  Therefore, we leverage a recently open-sourced 802.11ax MAC simulator \cite{lightsim-ax} as a starting point of our implementation and add significant extensions to arrive at our \textit{vanilla WiFi-}7 implementation. It mainly consists of the candidate \SI{320}{MHz} bandwidth of WiFi-7, fully functional UL- and DL-OFDMA, and frame aggregation features. In order to enable TI research community to conduct further experiments, we have open-sourced our simulator~\cite{ti_wifi}. In developing our custom simulator, we make a reasonable assumption that the channel conditions and received power for the chosen modulation and coding scheme (MCS-9) result in a negligible path loss.
The fixed MCS setup helps to minimize the impact of link adaptation, enabling us to isolate the effects of frame aggregation and jitter buffer on the signal reconstruction quality. This means that the PHY layer packet losses occur purely due to collisions. This is a common approach adopted by many existing works, such as \cite{naik}.
 The configuration details of the WiFi-7 network are given in Table~\ref{tab:wifi}. In this work, we transmit TI traffic using voice AC (AC\_VO) since it has the highest priority as per existing WiFi standards.

Accounting for several sensors and HD-quality video, we consider a scanerio where the UL (DL) traffic from (to) each STA is \SI{20}{Mbps}.
To reiterate, UL traffic is served each time a STA wins the contention as well as when AP schedules UL-OFDMA. On the other hand, the DL traffic is served only when AP schedules DL-OFDMA. While the UL and DL traffic are equal, the UL and DL access is asymmetric. Hence, in our implementation the AP schedules DL-OFDMA transmissions first. Only after the DL transmit queue is exhausted, UL-OFDMA is triggered. 
We use the standard \SI{1}{kHz} sampling of haptic updates. Each haptic sample elicits an MPDU. To leverage the periodic nature of MPDU generation in reducing channel collisions, we employ only SA mode for UL-OFDMA.  

For scheduling the STAs within UL- and DL-OFDMA, we use a simple approach for maximizing the bandwidth utilization, similar to the one recently proposed in \cite{magrin}. We employ a \textit{highest-first} approach where STAs with higher transmit queue occupancy are scheduled earlier than others. The AP sorts the STAs based on the decreasing order of their queue occupancy. It then divides them into disjoint groups of two if the number of STAs with data is lower than four, and groups of four otherwise. The channel bandwidth is then divided such that the number of RUs equals the group size. The group with the highest queue occupancy is served first. 
In case of DL-OFDMA, the AP simply looks at its transmit queue to determine the scheduling order of STAs. In case of UL-OFDMA, we leverage the periodic nature of MPDU generation to minimize the control overhead. The STAs inform about their queue occupancy to the AP by piggybacking the information in QoS control field of the packets -- this is known as \textit{unsolicited buffer status report}~\cite{ref:80211ax}.

\begin{figure}[!t]
    \centering
    \includegraphics[width=0.4\textwidth]{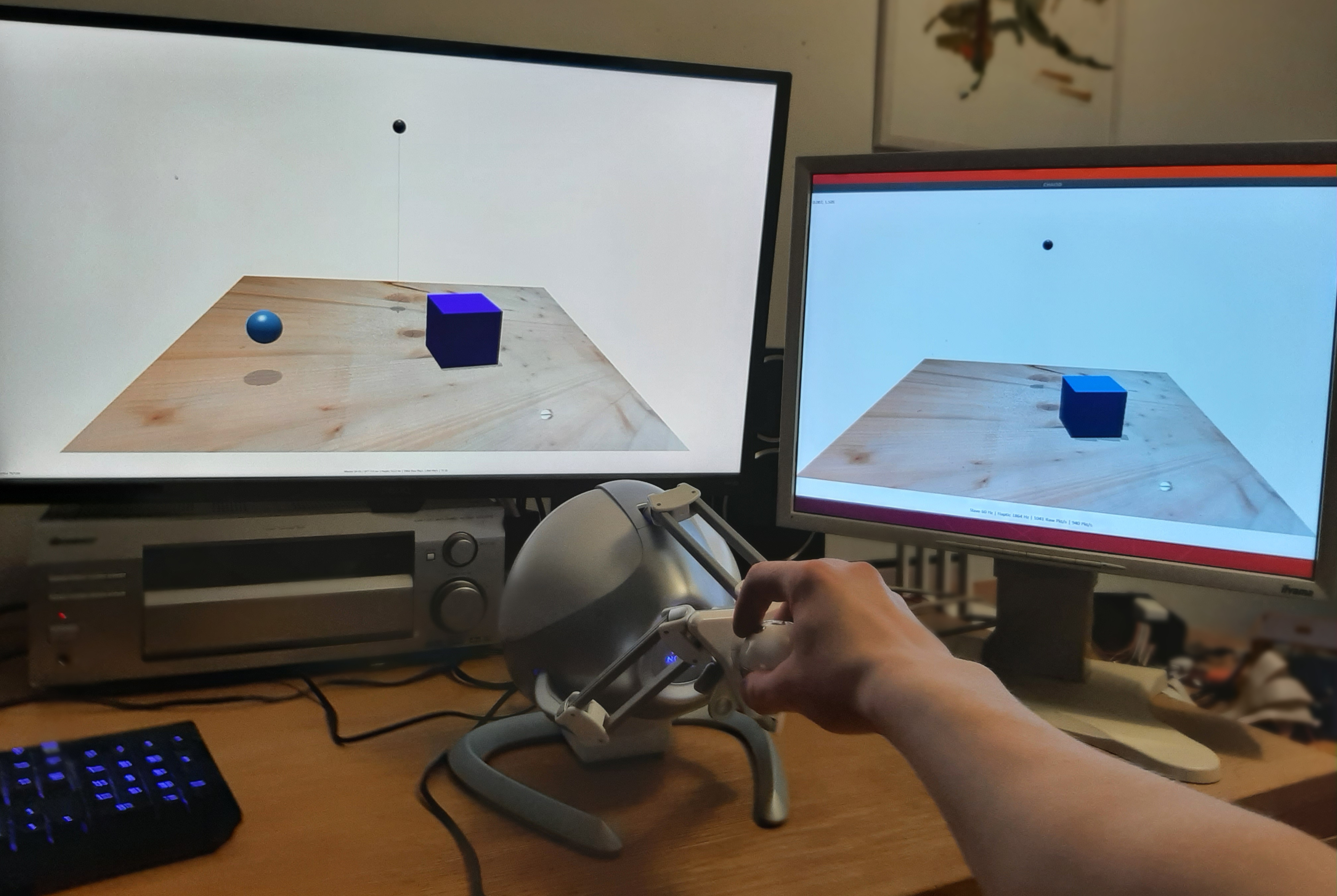}
    \caption{Networked VE game using a haptic device. The displays on the left and right belong to the master and controlled domains, respectively.}
    \label{fig:vegame}
\end{figure}

\begin{figure*}[!tbp]
\centering
  \begin{minipage}{0.48\textwidth}
    \includegraphics[width=0.98\textwidth]{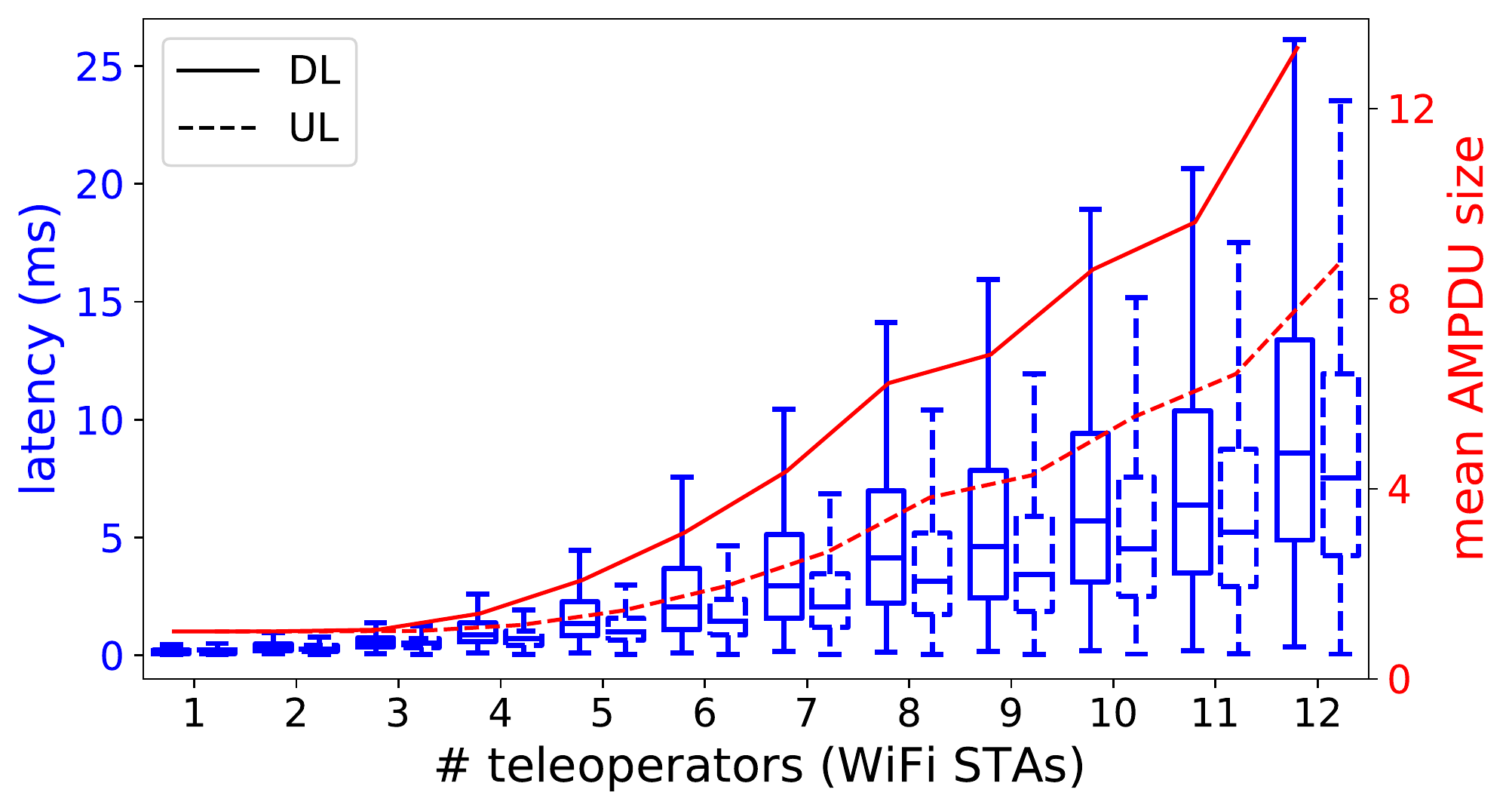} 
    \caption{Plot showing UL and DL latency and corresponding mean AMPDU size over a range of teleoperators (WiFi STAs) for vanilla WiFi-7 implementation.}
    \label{fig:lat_agg}
  \end{minipage} \hfill
  \begin{minipage}{0.48\textwidth}
    \includegraphics[width=0.98\textwidth]{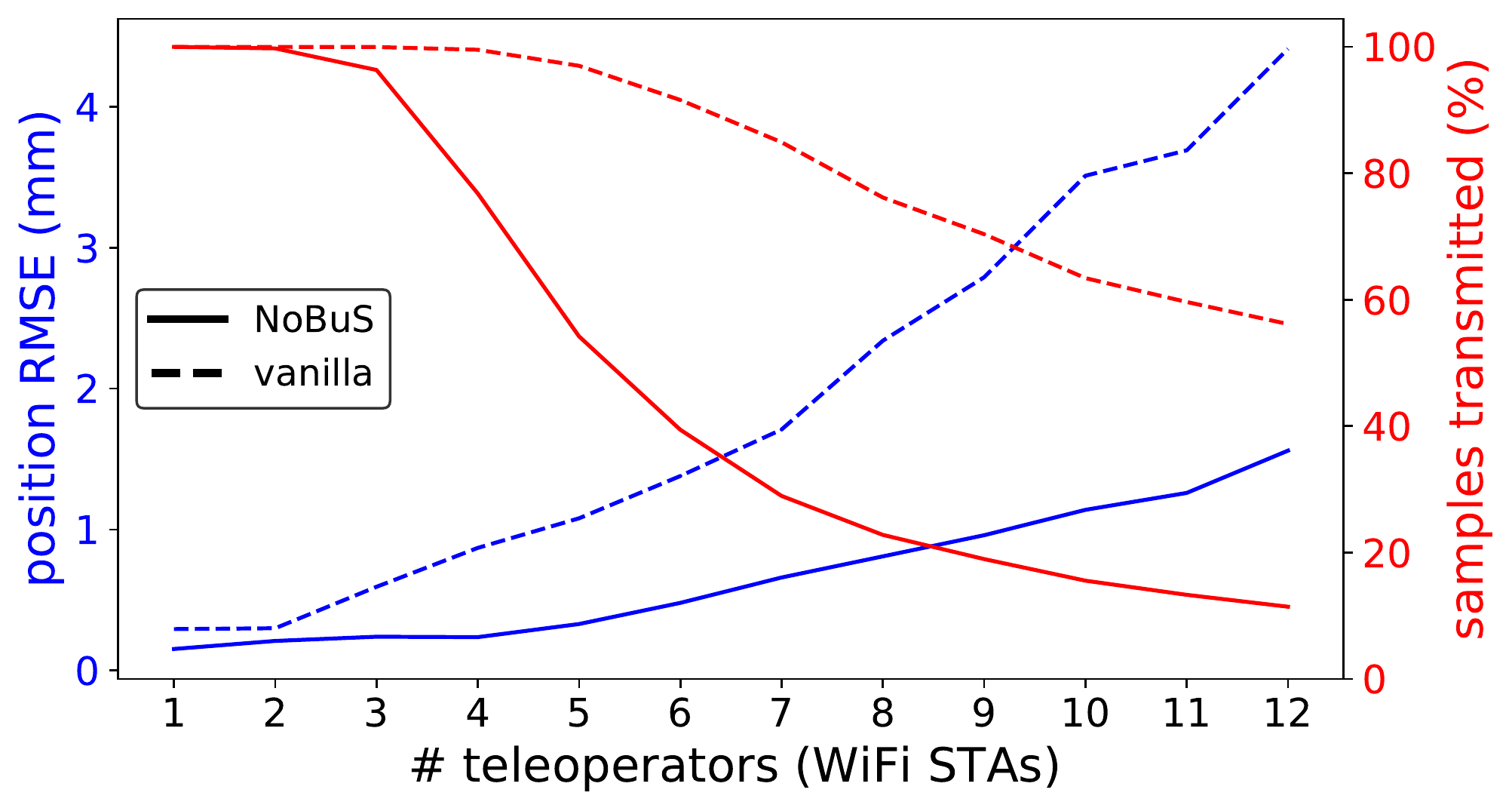} 
    \caption{Performance improvement of NoBuS in terms of position RMSE at teleoperator despite its lower fraction of transmitted samples in comparison with vanilla WiFi-7.}
    \label{fig:rmse-drop}
     
  \end{minipage}
\end{figure*}

\subsubsection{Realistic TI data generation}
\label{subsec:traces}
Since we intend to conduct performance assessment under realistic TI settings, we leverage a TI game that we developed in-house. In this, the human subject employs a Novint Falcon haptic device to move a rigid cube to a pre-determined target location inside a virtual environment (VE), as shown in Fig.~\ref{fig:vegame}. The VE runs on a remote server; further details on the VE setup are available in \cite{tixt}. Through human-in-the-loop experiments, we record the operator's motion commands, force feedback, and graphics data generated during this interaction. We then utilize the recorded data traces in our network simulator, where each operator/teleoperator uses samples from these traces as the payload. 
This helps us to conduct realistic network simulations while using data generated through an actual TI setup.

\subsection{Vanilla WiFi-7}
\label{subsec:perf}

\subsubsection{Packet generation}
\label{subsub:framevanilla}
In practice, the operator/teleoperator could have multiple sensors, each operating at a different sampling rate than the other. To minimize packetization delay under heterogeneous sampling rates, the application layer generates messages at the highest sampling rate of all sensors. A message carries the latest sample from each sensor and each message elicits an MPDU. Whenever a WiFi device gets medium access, it composes AMPDU by aggregating all MPDUs in its transmit queue corresponding to a destination. A packet is then transmitted consisting of an AMPDU. Essentially, in vanilla WiFi-7, each packet may consist of several samples from each sensor due to aggregation. The MAC transmit queue of vanilla WiFi-7 is depicted in Fig.~\ref{fig:factory}.

\subsubsection{Performance Evaluation}
\label{subsec:vanillaperf}
We begin by presenting UL and DL latency with increasing STA density (number of STAs) in Fig.~\ref{fig:lat_agg}. Note that since we are proposing an enhancement to WiFi-7, we measure only the latencies arising from WiFi layers -- MAC and PHY.
With twelve STAs, the worst-case DL and UL latencies are approximately \SI{26.10}{ms} and \SI{23.53}{ms}, respectively, resulting in a worst-case round-trip latency of \SI{49.63}{ms}. To get an idea of the extent of frame aggregation, Fig.~\ref{fig:lat_agg} also presents the mean AMPDU size which is the mean number of MPDUs per AMPDU. For low STA density, the AMPDU size is close to 1, indicating that the transmission rate can cope with the data generation rate without aggregation. With increasing STA density, the mean AMPDU size also starts to rise quickly and exceeds 12 and 8 on DL and UL, respectively, with twelve STAs. For \SI{1}{kHz} haptic sampling rate, this signifies that each AMPDU carries samples worth more than \SI{12}{ms} and \SI{8}{ms}, respectively, corresponding to each sensor. 

Note that the latency presented in Fig.~\ref{fig:lat_agg} correspond to only channel access and does not take into account the latency due jitter buffer at the receiver. While the earliest sample in an AMPDU is displayed first, the later samples are queued up in the jitter buffer for their turn to be displayed serially at \SI{1}{kHz}. Accounting for this, the round-trip latency is even higher than above specified values. Note that for TI applications, it is crucial to contain the worst-case round-trip latency within the ULL bound (\SI{10}{ms}). It can be seen that this deadline gets violated with as few as five STAs. Therefore, with vanilla WiFi-7, TI applications are extremely vulnerable to the detrimental effects of high latency.

\subsubsection{Contemplating frame aggregation}
\label{subsub:frameagg}
Since vanilla WiFi-7 maximizes the amount of samples delivered, the teleoperator can accurately replicate the operator's actions.
On the other hand, the jitter buffer latency which is the additional time to synchronize with the operator makes the teleoperator sluggish. From a control systems perspective, it is pragmatic to display only the latest sample. If the teleoperator was programmed to display only the latest received sample while discarding earlier ones, then the synchronization would be achieved in a significantly shorter amount of time. This approach, however, trades off the reliability of teleoperation for latency.

\section{Enabling Tactile Internet Over WiFi-7}
\label{sec:proposed}
In this section, we explore this latency-reliability trade-off through the design of a simple MAC layer enhancement that we call \textit{Non-Buffered Scheme (NoBuS)}.
We use the performance of vanilla WiFi-7 presented in Sec.~\ref{subsec:vanillaperf} as the baseline for gauging the improvement of NoBuS.

\subsection{Non-Buffered Scheme (NoBuS)}
\label{subsec:nobus}

\subsubsection{Working Principle}
In existing WiFi systems, when packet loss occurs MAC layer retransmits the samples up to the retransmission limit (Table~\ref{tab:wifi}). While attempting to retransmit the older samples increases latency (frame aggregation), the other extreme approach of completely disabling retransmissions would heavily deprive the teleoperator of even the latest samples. Hence, we require an in-between approach where the retransmission happens only until retransmission limit is reached or a new sample arrives -- whichever is earliest.  

Motivated by this, we design NoBuS as a simple MAC layer enhancement for enabling quick synchronization between the master and controlled domains. The working principle of NoBuS is as follows: When a device wins the channel contention, it transmits only the latest MPDU. This means that if there are multiple MPDUs buffered for the same destination, then all the previous MPDUs are discarded. Essentially, we use use sampling duration as the packet lifetime. This approach minimizes jitter buffer latency and transmission latency. Fig.~\ref{fig:factory} contrasts the MAC transmit queue of NoBuS with that of vanilla WiFi-7. Note that although we depict the MAC layer of AP alone, we implement NoBuS at the STAs also. 

\subsubsection{Packet generation}
The application layer messages are generated as explained in Sec.~\ref{subsub:framevanilla}.
Since NoBuS replaces the older MPDU in the transmit queue with the latest, each packet consists of a single MPDU carrying the latest sample from each sensor.

\subsection{Performance Evaluation}
Due to space constraints, we present the results for DL only while noting that the UL measurements follow similar trend.
As can be seen in Fig.~\ref{fig:rmse-drop}, NoBuS transmits a significantly lower amount of samples (up to \SI{50}{\%}) than vanilla WiFi-7. This is because NoBuS drops samples in two scenarios: i) exceeding retransmission limit, and ii) proactive dropping of older samples. On the other hand, vanilla WiFi-7 drops samples only in the former scenario. 

We now assess the impact of the above higher losses on the quality of signal reconstruction.
To this end, we leverage the recorded TI data traces (Sec.~\ref{subsec:traces}).
For the purpose of presentation, we consider only one axis of the 3D position data. The position coordinates in this axis vary predominantly in the range [-10, 10] cm. We use RMSE as the objective metric. Despite the higher losses, it can be seen that the RMSE of NoBuS increases only marginally with the STA density. Additionally, NoBuS yields a significantly lower RMSE than vanilla WiFi-7, with a maximum reduction of up to \SI{65}{\%}. Further, in TI applications involving very fast hand movements, we expect vanilla WiFi-7 to perform even worse and the NoBuS improvement to be higher. The reason behind this high performance of NoBuS is its ability to minimize the overall latency. As per our simulations, the worst-case DL and UL latency of NoBuS are around \SI{1.43}{ms} and \SI{2.29}{ms}, respectively, leading to a worst-case round-trip latency of \SI{3.72}{ms}. This is an order of magnitude lower than that of vanilla WiFi-7, and is also compliant to the necessary ULL limit of \SI{10}{ms} even in highly dense environments. This is because the worst-case queuing latency of a transmitted sample is only \SI{1}{ms}. An additional benefit of NoBuS is that the receiver's MAC layer can forward MPDUs to higher layers without waiting for the earlier ones. This avoids the latency arising from packet reordering.

From the standpoint of RMSE, the performance improvement of NoBuS reveals that  fast synchronization between operator and teleoperator is way more crucial for high quality TI communication than maximizing the number of transmitted samples. This aligns very well with the claims made in \cite{etvo} and hence highlights the need to focus on providing ULL service guarantees for TI applications rather than ultra-reliability. This opens up several opportunities and serves as an important insight for designing efficient TI systems.

Although we showed the efficacy of NoBuS in a WiFi network, the principle of \textit{non-buffering} can be explored on wired or other wireless networks where schemes akin to frame aggregation are typically employed. While we expect NoBuS' ULL guarantee to facilitate seamless teleoperation, further experimental verification through psychophysical studies is paramount for a systematic substantiation of the efficacy of NoBuS. 
We intend to take this up as a future extension of this work.

\subsection{Implementation details}
 Due to its simple working principle, the implementation of NoBuS on WiFi devices is quite straightforward. The MAC layer needs to be informed about the AC used by the TI stream. WiFi's default frame aggregation functionality should be disabled for that AC. At the STAs, the MAC layer can be configured to have a transmit queue of size 1 for the specific AC assigned for TI streams. The MAC layer should be programmed to simply replace the existing MPDU in the queue with the latest one. On the other hand, since the AP holds MPDUs for multiple STAs (for DL transmission), its queue is longer (ideally as long as the number of STAs). On arrival of an MPDU meant for DL transmission, the AP should identify the older MPDU to be replaced based on the destination MAC address. 
It is worth noting that NoBuS' operation does not entail any cross-layer designs.
Therefore, it does not violate any design principles of the WiFi standards and can be easily incorporated with the existing WiFi systems. Further, NoBuS places negligible additional computational and memory requirements on the WiFi devices. In our opinion, at some point, enhancements will be done to support TI traffic on WiFi, and NoBuS can be one of the submissions.

\section{Vistas for Future}
\label{sec:lessons}

\begin{itemize}[leftmargin=*]
    \item\textbf{Comprehensive evaluation}: 
    While the RMSE measurements show the potential of NoBuS to facilitate high-fidelity signal reconstruction, subjective experiments are important for further substantiation. 
    Another interesting research direction is to explore the applicability of \textit{age of information} theory to further enhance the performance of NoBuS.
    Also, we would like to demonstrate the working of NoBuS on a commercial WiFi chipset. This helps us to understand the performance of NoBuS under realistic wireless conditions. 
    \item \textbf{Scheduling framework}: Designing a transmission scheduling framework for haptic and video samples taking into consideration their temporal relationship is crucial. Further, doing so in presence of MLO and preemptive scheduling schemes is also a promising research avenue.
\end{itemize}

\section{Conclusions}
\label{sec:conclusions}
Tactile Internet (TI) will open umpteen possibilities for novel immersive applications involving haptic feedback. To this end, advancements in every hop of the network are a must. In this context, the first/last mile is expected to be invariably served by WiFi-7. In a first-of-its-kind, we considered the performance evaluation of WiFi-7 under realistic TI application conditions. 
We focused on the standard practice of using frame aggregation in conjunction with jitter buffer and experimentally demonstrated its negative impact on the quality of teleoperation. Through the design of \textit{NoBuS}, we showed remarkable improvement in latency and RMSE compared to vanilla WiFi-7. 
The promising findings of our work show potential for enabling TI communications over WiFi-7. 

\section*{Acknowledgment} 
This work has been undertaken in the \textit{Internet of Touch} project sponsored by Cognizant Technology Solutions and Rijksdienst voor Ondernemend Nederland under PPS O\&I.

\bibliographystyle{IEEEtran}
\bibliography{references}

\end{document}